\documentclass[12pt]{aastex}
\usepackage{pslatex}
\usepackage{verbatim}
\usepackage{graphicx}
\usepackage{amssymb}
\usepackage{lscape}
\usepackage{amsmath}
\usepackage{wrapfig}
\usepackage{float}
\usepackage{verbatim}
\usepackage{rotating}
\fontfamily{ptm}
\usepackage{hyperref}

\usepackage{pdfpages}

\usepackage{tocloft}
\usepackage{natbib}
\renewcommand{\icarus}{{\it Icarus}}

\newenvironment{itemize*}%
  {\begin{itemize}%
    \setlength{\itemsep}{3pt}%
    \setlength{\parskip}{0pt}}%
  {\end{itemize}}

\newenvironment{enumerate*}%
  {\begin{enumerate}%
    \setlength{\itemsep}{3pt}%
    \setlength{\parskip}{0pt}}%
  {\end{enumerate}}

\usepackage{color}
\usepackage{ulem}
\newcommand{\ACBc}[1]{\textcolor{black}{ #1}}

\begin{document}

\title{Overcoming the Meter Barrier and The Formation of Systems with Tightly-packed Inner Planets (STIPs)}
\author{A.~C.~Boley\altaffilmark{1},
M.~A.~Morris\altaffilmark{2}, and
E.~B.~Ford\altaffilmark{3}}

\altaffiltext{1}{Department of Physics and Astronomy, The University of British Columbia, 6224 Agricultural Rd., Vancouver, B.C.~V6T 1Z4, Canada}
\altaffiltext{2}{Center for Meteorite Studies, Arizona State University, P.O. Box 876004, Tempe, AZ 88287-6004, USA}
\altaffiltext{3}{Center for Exoplanets and Habitable Worlds, The Pennsylvania State University, 525 Davey Laboratory, University Park, PA 16802, USA; Department of Astronomy and Astrophysics, The Pennsylvania State University, 525 Davey Laboratory, University Park, PA 16802, USA}

\begin{abstract}
We present a solution to the long outstanding meter barrier problem in planet formation theory. 
As solids spiral inward due to aerodynamic drag, they will enter disk regions that are characterized by high
temperatures, densities, and pressures. 
High partial pressures of rock vapor can suppress solid evaporation, and promote collisions between partially molten solids, allowing rapid growth.
This process should be ubiquitous in planet-forming disks, which may be evidenced by the abundant class of Systems with
Tightly-packed Inner Planets (STIPs) discovered by the NASA {\it Kepler} mission. 
\end{abstract}

\maketitle

\section{INTRODUCTION}

The {\it Kepler} space mission has revealed numerous planetary types and systems, shaping our understanding of planet formation \citep{borucki_etal_apj_2011, batalha_etal_apjs_2013}.
Among the quickly-growing data is a subclass of multi-planet configurations referred to as Systems with Tightly-packed Inner Planets (STIPs). 
Their large abundance ($>10$\% of stars) suggests that they are one of the principal outcomes of planet formation. 
The prototype STIP is Kepler-11 \citep{lissauer_etal_nature_2011,lissauer_etal_arxiv_2013}, which hosts six known transiting planets, five of which have measured masses in the super-Earth and mini-Neptune regimes. 
The known planetary orbits in this system are spaced between $a=0.09$ and 0.47 AU, with small eccentricities and mutual inclinations. 

This dynamically cold configuration suggests that gravitational interactions between the planets were minimal during formation or that the disk was strongly dissipative.
The orbits are not in low-order mean-motion resonances, a key signature of  smooth disk migration, suggesting that migration may not have played a dominant role.
Although disk turbulence could be responsible for producing some STIPs that have planets near commensurabilities \citep{pierens_etal_aap_2011},
 {\it in situ} formation seems to be the simplest solution for most STIPs \citep[for a summary see][]{raymond_etal_mnras_2008}.
Such formation would require growth of massive planets on the stellar side of the water ice line, which is difficult to reconcile with the current planet formation paradigm. 
 Nonetheless, we must entertain the idea that the current paradigm is incomplete and that  super-Earth and mini-Neptune formation at short orbital periods is plausible  \citep{chiang_laughlin_mnras_2013}.
This requires the delivery and retention of significant material into the inner nebula.

In this Letter, we demonstrate an overlooked mechanism that should be prevalent in \ACBc{many} planet-forming disks, lead to planet formation at high disk temperatures, and overcome the meter barrier.

\subsection{The Meter Barrier}

The ``meter barrier'' is the difficulty in gradually forming planetesimals from small solids, as aerodynamic forces will cause rapid migration of rocks and boulder-sized solids ($\sim 10$-100 cm) in a nearly Keplerian disk \citep{adachi_etal_1976,weidenschilling_mnras_1977}.
Because the disk has a pressure gradient, a monotonically decreasing pressure will cause the azimuthal speed of the gas to always be less than the Keplerian orbital speed $v_K$.
A solid, which does not have pressure support and moves at $v_K$, will thus orbit with a head wind.  
This causes an exchange of angular momentum, and the solid spirals inward.  
For very large particle sizes, the stopping time ($t_s$) is  very long compared with the orbital period at 1 AU, preventing rapid in-spiral.  
For very small particles, $t_s$ is very short, but the terminal radial velocity is also very small, again preventing rapid in-spiral. 
Whenever the term $t_s v_K/\varpi = t_s \Omega_K =\tau\sim 1$, for disk radial distance $\varpi$ and Keplerian orbital frequency $\Omega_K$, very efficient inward drift occurs. 
Taking rough values for $\varpi=1$ AU in an envisaged planet-forming disk ($\rho\sim 10^{-9}$ g/cc and $T\sim 300$ K), the most rapid drift size corresponds to about 1 m, although notable drift will begin in the mm to cm-size regime.
For pressure gradients in typical disk models, the in-spiral time for a meter-sized object at 1 AU is only a few 100 yr, much shorter than the timescale to form a planet at this location \citep[see, e.g.,][]{weidenschilling_mnras_1977}.  
This appears to inhibit planet formation and has been named the ``meter barrier problem.''
At both much smaller and larger sizes, the in-spiral time becomes long relative to 1 m.

An additional component of the meter barrier problem is that high relative particle speeds, due to inward drift and/or turbulence, will be destructive as sizes approach $\sim 10$ to 100 cm.  
This is emphasized in  \cite{blum_wurm_review_2008} (see their Fig.~12). 
Even if large solids could be preserved at their given location by turbulence, growth beyond about 10 cm becomes inhibited. 
Previously proposed solutions include gravitational collapse of the solids due to concentrations in, for example, streaming instabilities \citep{youdin_goodman_2005, johansen_etal_apjl_2009, bai_stone_apj_2010}.

We are thus left with two basic solutions to the meter barrier problem:  Either (1) rapid formation of planetesimals must occur through secondary instabilities,  or (2) the collisional process and inherent outcome of large solid interactions bust be modified.   The work here explores the second option, which we argue represents a fundamental process in disk evolution and planet formation.  We further stress that the model presented here is not intended to operate {\it instead} of, e.g., a streaming instability (option 1), but represents a different pathway that may even promote instabilities.

\section{Suppression of Rock Evaporation in the Inner Regions of Planet-Forming Disks}

\ACBc{The local mass density of solids can become very large in the inner regions of disks, even with a fixed solid-to-gas ratio.  This will promote an increased solid collision rate, which could be further enhanced if any solid concentrations were to occur (e.g., some midplane settling).
Nonetheless, solids are expected to evaporate (either from liquidus or solidus) at the high temperatures of the inner disk, causing many solids to be lost in this high-density region.}
 The exact boundary depends on several factors, but it is often assumed that solids will evaporate at temperatures in excess of 1400-1500 K.
To determine whether particles are lost to evaporation, it is necessary to take into account the saturation vapor pressure of rock and the {\it timescale} for the evaporation of solid material relative to collision times as applicable to conditions in disks at radii $\sim 0.1$ AU.
 
Whenever the partial pressure of a vapor is equal to its saturation pressure, evaporation and re-condensation will be in equilibrium.  
For a gas with only a single species, the evaporation rate \ACBc{(mole per area per time)} and saturated vapor pressure are related by the Hertz-Knudsen relation:
\begin{equation}
 J=\frac{\gamma P_s}{\left(2\pi m R T\right)^{1/2}},
\end{equation}
where $m$ is the weight of the species, $T$ is the gas temperature with gas constant $R$, $\gamma$ is the evaporation coefficient, and $P_s$ is the saturation vapor pressure \citep[e.g.,][]{richter_etal_2002_geocosmo}.  
As a proxy for evaporation of rock, we focus on Mg in the context of forsterite (Mg$_2$SiO$_4$), an abundant mineral in chondrites.  
Following \cite{richter_etal_2002_geocosmo}, the evaporation rate is
\begin{equation}
J\approx J_0 \exp(-E_0/(RT)) \left(P_{\rm H_2}/P_0\right)^{1/2}
\end{equation}
The constants are set to match laboratory experiments at different temperatures and H$_2$ pressures $P_{\rm H_2}$, which we calibrate here using the figures and results presented in \cite{richter_etal_2002_geocosmo} for Mg.
 Thus, $E_0=300$ kJ mole$^{-1}$,  $J_0 = 75$  mole cm$^{-2}$ s$^{-1}$ with $P_0=1$ mbar. 
The evaporation coefficient is approximated by $\gamma\sim150\exp(-15000{\rm~K}/T)$.

Setting $P_{\rm H_2}\sim 10$ mbar, $T\sim1500$ K,  and the  volume mixing ratio $r\sim 3.8\times10^{-5}$ \citep[solar;][]{lodders_2003}\footnote{We use a protosolar value $A=7.62$, and assume a hydrogen number density  0.92 the total number density.}, we find \ACBc{$P_{\rm Mg}/P_s\sim0.1$} if the local Mg abundance is entirely vapor.
The implication is that, in a  solar composition nebula, a solid enhancement \ACBc{greater than about 10} relative to bulk  composition will prevent rocks from evaporating \ACBc{for the given temperature and pressure}.

\begin{figure}[h]
\includegraphics[width=4in,angle=-90]{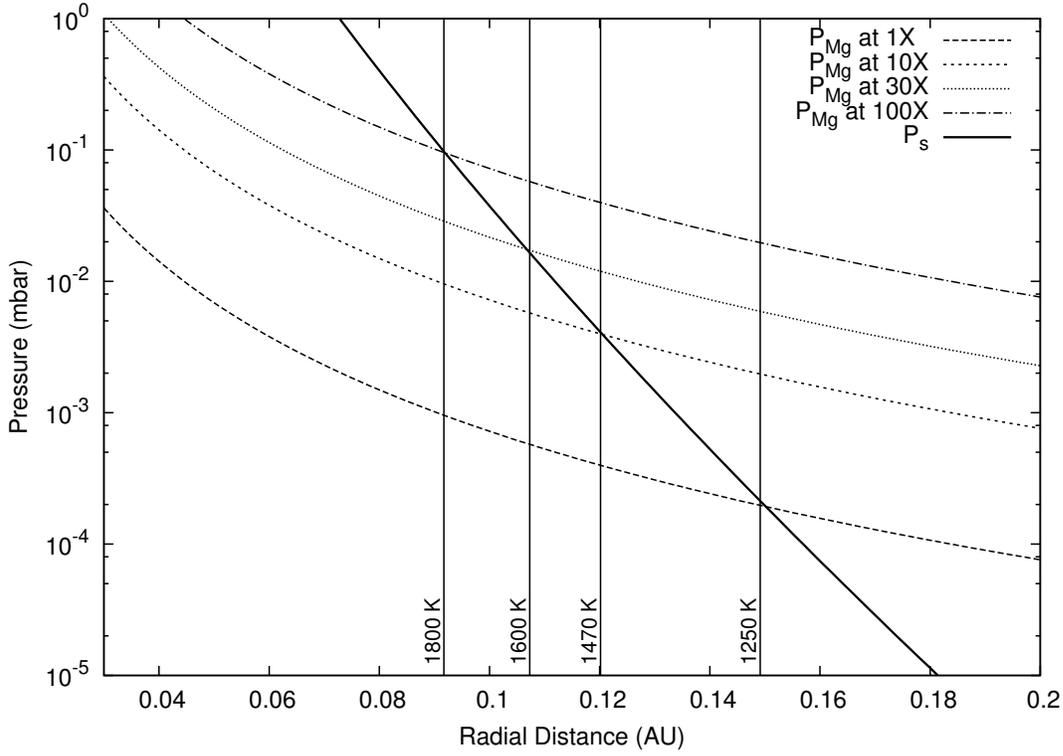}
\caption{ Saturation vapor pressure of Mg (solid curve) compared with different hypothetical partial pressures (dashed/dotted curves). 
The solar volume mixing ratio for Mg is taken to be $3.8\times10^{-5}$.  Higher or lower values could be due to differences in metallicity or degree of concentration/depletion of solids.
When the partial pressure is above the saturation pressure, net evaporation becomes suppressed;
large volume mixing ratio enhancements can suppress net evaporation at 1500 K. 
\ACBc{Locally enhanced partial pressures due to self-shielding effects (vapor clouds) could also lead to equilibration}.
\label{fig:evap}}
\end{figure}

Consider a simple disk model that has a radial temperature profile characterized by $T\left(\varpi\right)=300 \left(\frac{\varpi}{\rm AU}\right)^{-3/4}$ K and total gas-mass density $\rho\left(\varpi\right)=10^{-9}\left(\frac{\varpi}{\rm AU}\right)^{-2.5}$ g/cc.  
These values are only meant to be illustrative, and variations are possible without changing the basic properties discussed here. 
Figure \ref{fig:evap} shows the Mg pressure profile of the resulting disk for four different volume mixing ratios of a hypothetical gas, assuming all Mg is in vapor.
The 1X curve is based on the solar mixing ratio of Mg, an ideal gas, and a mean molecular weight $\mu=2.3$.
Curves represented by 10X, etc., are 10X the Mg partial pressure at solar abundance ($\mu$ will be different in these cases).  
For a solar abundance nebula, if all the Mg were in vapor, the partial pressure would become much less than the saturation pressure for temperatures above $\sim 1250$ K.  
For an enhancement of Mg vapor that is ten times the solar abundance (10X), evaporation is suppressed until about 1470 K.  
At even higher concentrations of 30Z and 100Z, evaporation will be suppressed until temperatures of 1600 and 1800 K are reached, respectively.   
For high volume mixing ratios, which may be obtained through high initial metal abundances and/or inward drift of solids, net evaporation can be suppressed.

 The above limit assumes that newly-formed vapor can diffuse away from an evaporating solid instantaneously.  
As pointed out by \cite{cuzzi_alexander_2006}  the overlapping effects of evaporating solids can allow ensembles of solids/droplets to equilibrate with their own collective  vapor.  While their work was done in the context of chondrule formation (unlike this study), the basic physics is general and applicable to our model for the rapid growth of solids in the inner nebula.
Net evaporation can be suppressed \ACBc{whenever $\exp(-n \pi s^2 \gamma v_{\rm th} t)=\exp(-\xi)\sim 0$} \citep[see][]{cuzzi_alexander_2006}, where $s$ is some representative size for solids that have a number density $n$, $v_{\rm th}$ is the thermal velocity of the vapor, and $\gamma$ is the evaporation coefficient.   
The time $t$ is a characteristic time for the problem, which we take to be the nominal evaporation timescale
\begin{equation}
t^{\rm ev}\approx 2 \rho_m s/(3 J~140~\rm g~mole^{-1}).\label{eqn:evap}
\end{equation}
The factor of two takes into account that two Mg must be lost for every forsterite.
A  1 $\micron$ grain  with $\rho_m=2.5$ g/cc at 1500 K will evaporate in about 2 minutes at $P_{H_2}\sim 10$ mbar, where this H$_2$ pressure is meant to be illustrative of the types of conditions that can be found at 0.1 AU around a solar-mass star.
In contrast, a cm-sized grain requires \ACBc{$10-20$} days to evaporate under these conditions, or  $\sim1$-2 orbital periods at 0.1 AU.

Now consider the case where all the mass is in solids of size $s$.  
\ACBc{In this case,  $\xi=\gamma \frac{3 \rho_s}{4 \rho_m}\frac{ v_{\rm th}}{s} t^{\rm ev}$.  
To allow evaporation and re-condensation to equilibrate, $\xi\gtrsim3{\rm -}6$, which corresponds to $\rho_s\gtrsim  1{\rm- }2\times10^{-8}$ g/cc ($\gamma\sim 0.007$ at 1500K).  
However, equilibration is not strictly necessary, as we only require the effective evaporation time to be much longer than the coagulation timescale.  For $\xi\sim1$, the effective evaporation time will be increased by a factor of few over the nominal $t^{\rm ev}$, with a corresponding $\rho_s\gtrsim  2{\rm- }3\times10^{-9}$ g/cc. 
A reasonable estimate of the total mass fraction of refractory material that is drifting into radii $\varpi\sim0.1$ AU is approximately $0.00375$ (\ACBc{i.e., for solar metallicity and no additional concentration}). Accordingly, in our envisaged disk $\rho_{s}\sim 10^{-9} \left(0.1\right)^{-2.5} 0.00375\rm{~g/cc}\sim 10^{-9}$ g/cc. 
Even if the far-field partial pressure does not exceed the saturation pressure at $\sim 0.1$ AU, inward drifting 1-cm solids will begin to show notable self-shielding effects at concentrations of a few times solar, and suppression of net evaporation could occur for concentrations $\gtrsim10$-20 solar.
This estimate does not strongly depend on the actual representative grain size due to the direct dependence of $t^{\rm ev}$ on $s$.}

The above situation becomes sustainable if the evaporation front  is a few times $\sim\sqrt{D t^{\rm ev}}$, i.e., the distance vapor diffuses away from the ensemble of solids during evaporation. 
The diffusion coefficient $D\sim \frac{0.00014}{P({\rm bar})} T({\rm K})^{1.5}$ cm$^2$ s$^{-1}$ sets the rate at which vapor can diffuse through H$_2$ gas.  
 \ACBc{For our envisaged conditions, $D\sim 800$ cm$^{2}$ s$^{-1}$, requiring the evaporation front to be $L\gg$ 1 km to mitigate the effects of diffusion. 
The gas scale height $H$ at 0.1 AU  is about 400,000 km in our envisaged disk.  Thus, the evaporation front only needs to extend over a small fraction of $H$.  Solids will evaporate over a vertical distance that is comparable to the solid vertical scale height $H_s$, which we take to be comparable to the overall size of $L$.   The ratio $H_s/H$ is roughly the inverse of the concentration of solids in the midplane due to settling (see $K$ defined next section).  Even for very high midplane concentrations, self-shielding effects play a role in limiting the net evaporation rate of solids.   We also need to consider whether the solids themselves can move out of their own vapor cloud due to radial drift.  Assuming that cm-sized solids migrate inward between about $v_{\rm drift}\sim10$ and 100 cm/s, the radial dimension of the evaporation front can extend for at least $\sim v_{\rm drift}t^{\rm ev}\sim 100$-2000 km. 
However, inward moving solids will always produce a collective vapor trail that will be seen by solids entering the evaporation front, and diffusion in the vertical direction may still be the most limiting condition.}

\section{Collisions in the Inner Regions of Planet-Forming Disks}

The calculations presented here are based on the evaporation of Mg.  
However, Si is expected to have a similar behavior  \citep{richter_etal_2002_geocosmo}.
While the detailed rates will change when considering bulk compositions of rocks, the overall picture should remain valid: rocks will not necessarily be destroyed by evaporation in the inner nebula.  Moreover, collisions will be very frequent in this environment, and the solids may be partially molten (discussed more below). 

The mass growth rate for a solid of size $s$ colliding with solids of similar or smaller sizes is $\dot{m} = \rho_s \pi s^2 v_{\rm vel}$ for local solid mass density $\rho_s$, cross section $\pi s^2$, and relative velocity $v_{\rm rel}$.  
The mass growth can be related to size growth by $\dot{s} = \frac {\rho_s}{\rho_m}\frac{ v_{\rm vel}}{4}$ for particle internal density $\rho_m$.
Both the solid mass density and the relative velocity will depend on the solid size and the degree of turbulence in the disk, which can be described using the $\alpha$ formalism \citep{cuzzi_hogan_2003}.
Following \citep{dubrulle_etal_1995},  $v_{\rm rel}=\alpha^{1/2} c \frac{\sqrt{2\tau}}{1+\tau}$ for sound speed $c$ ($\sim 3$ km/s at 0.1 AU).  
The amount of settling, and hence midplane concentration, $K = \left(\frac{\tau}{\alpha}\right)^{1/2}\left(1+\frac{\alpha}{\tau}\right)^{1/2}$, where $\rho_s = K \rho_{0,s}$.
Combining these relations, $\dot{s} = \frac{\rho_{0,s}}{\rho_m}\frac{c}{2\sqrt{2}}\frac{\tau}{1+\tau}\left(1+\frac{\alpha}{\tau}\right)^{1/2}$.
When $\tau$ is large, the growth approaches the Safronov limit.  At small $\tau$, radial drift by larger solids and Brownian motion will prevent the growth rate from becoming trivially small.  \ACBc{Note that the growth rate becomes nearly independent of $\alpha$, except at large $\alpha/\tau$. }

As done for the $\rho_s$ calculation above, we take $\rho_{0,s}\sim 10^{-9}$ g/cc for our envisaged disk. 
At these distances, sizes $s\sim 10$ cm have $\tau\sim 1$ \citep[based on][but for conditions at 0.1 AU]{weidenschilling_mnras_1977}.   
As particles approach this size, they will grow from collisions at about 2 cm/day, which will increase to about 4 cm/day as $\tau$ becomes large. 
We can thus expect 20-40 cm of growth per orbit.  The peak radial drift timescale is of order 100 local orbits, allowing the solid to grow above $\tau\sim 100$. 
The actual radial drift rate will decay rapidly during this time, creating a stable and limiting situation for growth. 
Higher metallicities or local concentrations of solids will enhance this effect.

\ACBc{These results are in reasonable agreement with existing literature. \cite{birnstiel_etal_2010} show that the meter barrier can be overcome for perfect sticking and a drift rate efficiency of 0.75 for growth at $\varpi\sim0.2$ AU in their model.  The growth rate per orbit will be more than twice as fast at 0.1 AU as it is at 0.2 AU for reasonable disk models, and no reduction of drift is necessary.   Moreover, as discussed next, our results physically motivate the assumption of perfect or near-perfect sticking, which the authors assumed as a test case. }

The assumption that collisions lead to growth is reasonable if kinetic energy can be dissipated during collisional growth.  
If solids are partially molten, then their viscosity may provide this dissipation (below).
\cite{ebel_grossman_2000} showed that silicate melts will be stable for a range of high temperatures (including 1500 K) at total pressures of 1 mbar and at rock vapor enhancements of about 100 solar.  
Fractional melts of say 10\% will require less extreme conditions.  
These experiments were only conducted at 1 mbar, while 10 mbar is more representative of the pressure in the inner nebula.  
The stability of melts is dependent on the pressure, which will reduce the necessary vapor enhancement further.  
For these reasons, the enhancement of 100 should be taken as an upper limit, but such a value may be attainable.

\cite{ciesla_2006_liquid}, motivated by compound chondrules, showed that hard spheres with $s\sim 0.3$ mm and a thin viscous surface layer can survive collisions $\sim 100$ m/s (viscosity $\eta\sim100$ poise).  
At larger speeds or larger solids, the thin surface layer will not dissipate all of the kinetic energy without a comparable increase in viscosity.
In the model explored by Ciesla, failure to dissipate kinetic energy did not by itself mean the solids were destroyed, rather, the model could no longer predict the outcome.  
However, a free-floating rock or boulder that is just experiencing melt may be better described as a solid suspension. 
 If the suspension is characterized by densely packed particles surrounded by layers of melted rock, then the rheology can be non-Newtonian and the effective viscosity  can become many orders of magnitude larger than the fluid's viscosity in isolation \citep[e.g.,][]{stickel_powell_2005}.  
For example, if the particles in the suspension are hard spheres with a volume filling ratio of $\phi$, with some maximum possible volume ratio $\phi_m\sim 0.6$, then the effective viscosity $\eta'=\eta\left(1+\frac{5\phi}{1-\phi/\phi_m}\right)^2$ as derived experimentally.  
Large viscosities could dissipate significant kinetic energy, including collisions in excess of 100 m/s.  A rough estimate of this dissipation is $E_{\nu}\sim 4\pi \eta v s^2/3$.  
 Comparing this with the kinetic energy gives $E_{\nu}/E_k \sim 2 \eta /(\rho_m s v)$.  
For a rock with $s\sim 100$ cm to survive a collision at $v\sim 100$ m/s with a comparable impactor requires $\eta\sim1$ to 2  Mpoise.  
This is a factor of $10^4$ larger than \ACBc{what may be typical for molten material alone}, but is feasible for a suspension where $\phi\rightarrow\phi_m$.
At small $\phi$, collisions of large rocks will not necessarily lead to growth, but because collisions should be taking place continuously, growth may begin at the onset of the initial melt stages.  

\ACBc{  The growth from planetesimals to planets is harder to estimate.  In particular, if the typical relative velocity for collisions becomes too high, the effective viscosities required to dissipate sufficient collisional energy may no longer be attainable.  These difficulties may be overcome, at least in part, by the highly dissipative and high density environment of the inner nebula, but this remains a topic for further study and will be needed to test the full viability of this model.}





\section{Discussion}

\ACBc{We have identified one potential solution to the outstanding meter barrier problem. If correct,  the process should be common in planet-forming disks and can lead to {\it in situ} formation of planets at short orbital periods. The mechanism has two components, both resulting from the environment of the inner nebula: (1)  Net evaporation from solid surfaces can become suppressed by high partial pressures, partly due to self-shielding effects.  (2) Collisional growth rates will be very large at short orbital periods as long as collisional destruction can be mitigated, which we suggest can be facilitated by collisions between partially molten solids.  In this context, solids  migrate to small disk radii, experience some melting without complete evaporation, and grow beyond the fast radial drift regime.   This mechanism does not require additional instabilities or radially localized total pressure enhancements, although such conditions would aid this process.}

\ACBc{While the above processes may be very common, additional work is required to understand the diversity of planetary system architectures. For example, the fraction of stars initially hosting STIP-like systems may be greater than that observed for main sequence stars due to subsequent orbital evolution sculpting the population of long-term stable planetary systems observed by {\it Kepler}. It is also possible that disk processes lead to bifurcations of planetary system architectures early in the formation process, with subsequent evolution playing a moderate role. }

The authors thank Conel Alexander and Fred Ciesla for their invaluable comments.  We also thank the referees for their helpful comments and suggestions.
ACB's contribution was supported by The University of British Columbia.  
MAM thanks Erik Asphaug and the Center for Meteorite Studies at Arizona State University for support.
EBF acknowledges support from the Penn State Center for Exoplanets and Habitable Worlds and a NASA Kepler Participating Scientist Program award (NNX12AF73G).

\bibliographystyle{apj}

\begin{thebibliography}{}

\bibitem[\protect\citeauthoryear{{Adachi}, {Hayashi}, \& {Nakazawa}}{{Adachi}
  et~al.}{1976}]{adachi_etal_1976}
{Adachi}, I., {Hayashi}, C.,  \& {Nakazawa}, K. 1976, Progress of Theoretical
  Physics, 56, 1756

\bibitem[\protect\citeauthoryear{{Asphaug}, {Jutzi}, \& {Movshovitz}}{{Asphaug}
  et~al.}{2011}]{asphaug_etal_2011}
{Asphaug}, E., {Jutzi}, M.,  \& {Movshovitz}, N. 2011, Earth and Planetary
  Science Letters, 308, 369

\bibitem[\protect\citeauthoryear{{Bai} \& {Stone}}{{Bai} \&
  {Stone}}{2010}]{bai_stone_apj_2010}
{Bai}, X.-N.,  \& {Stone}, J.~M. 2010, ApJ, 722, 1437

\bibitem[\protect\citeauthoryear{{Batalha} et~al.}{{Batalha}
  et~al.}{2013}]{batalha_etal_apjs_2013}
{Batalha}, N.~M., et~al. 2013, ApJS, 204, 24

\bibitem[\protect\citeauthoryear{{Birnstiel}, {Dullemond}, \&
  {Brauer}}{{Birnstiel} et~al.}{2010}]{birnstiel_etal_2010}
{Birnstiel}, T., {Dullemond}, C.~P.,  \& {Brauer}, F. 2010, \aap, 513, A79

\bibitem[\protect\citeauthoryear{{Blum} \& {Wurm}}{{Blum} \&
  {Wurm}}{2008}]{blum_wurm_review_2008}
{Blum}, J.,  \& {Wurm}, G. 2008, \araa, 46, 21

\bibitem[\protect\citeauthoryear{{Borucki} et~al.}{{Borucki}
  et~al.}{2011}]{borucki_etal_apj_2011}
{Borucki}, W.~J., et~al. 2011, ApJ, 736, 19

\bibitem[\protect\citeauthoryear{{Chiang} \& {Laughlin}}{{Chiang} \&
  {Laughlin}}{2013}]{chiang_laughlin_mnras_2013}
{Chiang}, E.,  \& {Laughlin}, G. 2013, MNRAS, 431, 3444

\bibitem[\protect\citeauthoryear{{Ciesla}}{{Ciesla}}{2006}]{ciesla_2006_liquid}
{Ciesla}, F.~J. 2006, Meteoritics and Planetary Science, 41, 1347

\bibitem[\protect\citeauthoryear{{Cuzzi} \& {Alexander}}{{Cuzzi} \&
  {Alexander}}{2006}]{cuzzi_alexander_2006}
{Cuzzi}, J.~N.,  \& {Alexander}, C.~M.~O. 2006, \nat, 441, 483

\bibitem[\protect\citeauthoryear{{Cuzzi} \& {Hogan}}{{Cuzzi} \&
  {Hogan}}{2003}]{cuzzi_hogan_2003}
{Cuzzi}, J.~N.,  \& {Hogan}, R.~C. 2003, \icarus, 164, 127

\bibitem[\protect\citeauthoryear{{Dubrulle}, {Morfill}, \&
  {Sterzik}}{{Dubrulle} et~al.}{1995}]{dubrulle_etal_1995}
{Dubrulle}, B., {Morfill}, G.,  \& {Sterzik}, M. 1995, \icarus, 114, 237

\bibitem[\protect\citeauthoryear{{Ebel} \& {Grossman}}{{Ebel} \&
  {Grossman}}{2000}]{ebel_grossman_2000}
{Ebel}, D.~S.,  \& {Grossman}, L. 2000, \gca, 64, 339

\bibitem[\protect\citeauthoryear{{Ida}, {Guillot}, \& {Morbidelli}}{{Ida}
  et~al.}{2008}]{ida_etal_2008}
{Ida}, S., {Guillot}, T.,  \& {Morbidelli}, A. 2008, \apj, 686, 1292

\bibitem[\protect\citeauthoryear{{Johansen}, {Youdin}, \& {Mac Low}}{{Johansen}
  et~al.}{2009}]{johansen_etal_apjl_2009}
{Johansen}, A., {Youdin}, A.,  \& {Mac Low}, M.-M. 2009, ApJ, 704, L75

\bibitem[\protect\citeauthoryear{{Lissauer} et~al.}{{Lissauer}
  et~al.}{2011}]{lissauer_etal_nature_2011}
{Lissauer}, J.~J., et~al. 2011, Nature, 470, 53

\bibitem[\protect\citeauthoryear{{Lissauer} et~al.}{{Lissauer}
  et~al.}{2013}]{lissauer_etal_arxiv_2013}
{Lissauer}, J.~J., et~al. 2013, ArXiv e-prints

\bibitem[\protect\citeauthoryear{{Lodders}}{{Lodders}}{2003}]{lodders_2003}
{Lodders}, K. 2003, \apj, 591, 1220

\bibitem[\protect\citeauthoryear{{Pierens}, {Baruteau}, \& {Hersant}}{{Pierens}
  et~al.}{2011}]{pierens_etal_aap_2011}
{Pierens}, A., {Baruteau}, C.,  \& {Hersant}, F. 2011, A\&A, 531, A5

\bibitem[\protect\citeauthoryear{{Raymond}, {Barnes}, \& {Mandell}}{{Raymond}
  et~al.}{2008}]{raymond_etal_mnras_2008}
{Raymond}, S.~N., {Barnes}, R.,  \& {Mandell}, A.~M. 2008, MNRAS, 384, 663

\bibitem[\protect\citeauthoryear{{Richter} et~al.}{{Richter}
  et~al.}{2002}]{richter_etal_2002_geocosmo}
{Richter}, F.~M., {Davis}, A.~M., {Ebel}, D.~S.,  \& {Hashimoto}, A. 2002,
  \gca, 66, 521

\bibitem[\protect\citeauthoryear{{Stickel} \& {Powell}}{{Stickel} \&
  {Powell}}{2005}]{stickel_powell_2005}
{Stickel}, J.~J.,  \& {Powell}, R.~L. 2005, Annual Review of Fluid Mechanics,
  37, 129

\bibitem[\protect\citeauthoryear{{Weidenschilling}}{{Weidenschilling}}{1977}]{weidenschilling_mnras_1977}
{Weidenschilling}, S.~J. 1977, MNRAS, 180, 57

\bibitem[\protect\citeauthoryear{{Youdin} \& {Goodman}}{{Youdin} \&
  {Goodman}}{2005}]{youdin_goodman_2005}
{Youdin}, A.~N.,  \& {Goodman}, J. 2005, ApJ, 620, 459

\end{thebibliography}

\end{document}